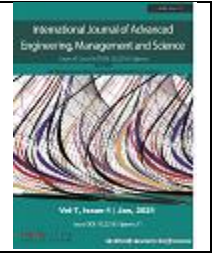

# Understanding the Farmers' Environmental Citizenship Behaviors Towards Climate Change: The Moderating Mediating Role of Environmental Knowledge and Ascribed Responsibility

Immaculate Maumoh*, Emmanuel H Yindi

Department of Public Policy, University of Science and Technology of China, Anhui-Hefei, China
*Corresponding Author

*Abstract*— *Knowledge is known to be a pre-condition for an individual's behavior. For the most efficient informational strategies for education, it is essential that we identify the types of knowledge that promote behavior effectively and investigate their structure. The purpose of this paper is therefore to examine the factors that affect Kenyan farmers' environmental citizenship behavior (ECB) in the context of Adaptation and mitigation (Climate smart agriculture). To achieve this objective, a theoretical framework has been developed based on value-belief-norm (VBN) theory. Design/methodology/approach – Data were obtained from 350 farmers using a survey method. Partial lease square structural equation modelling (PLS-SEM) was used to examine the hypothetical model. The results of PLS analysis confirm the direct and mediating effect of the causal sequences of the variables in the VBN model. The moderating role of Environmental knowledge has been seen to be impactful in Climate Smart Agriculture.*

*Keywords*— *Climate Change, Citizenship Responsibility, Environmental behavior, Environmental Knowledge, Local Ecological Knowledge.*

## I. INTRODUCTION

African countries have been facing enormous challenges in adapting to climate change as well as responding to the slowdown in economic growth in recent years. These twin crises have seen increased their being vulnerable and subsequently the effects of climate change and climate variabiity (Recha, 2019) . Sub-saharan Africa is heavily dependent on Agriculture and must adopt to the variances caused by variable climatic conditions. Kenyan households that engage in agriculture solely contributed 31.4% to the rural areas poverty reduction, and, agriculture stays as the largest income source for both poverty stricken and other households in rural areas, according to the most recent World Bank economic analysis (Ayeri, Christian, Josef, & Michael, 2012). The challenges that farmers face are sadly not only the direct effects of climatic variability but also the international demands such as food security , reduction of GHG emissions and many others (Bryan, Ringler, Okoba, Koo, et al., 2013). However, implementing changes is no mean feat due to numerous barriers (Chua et al., 2019 & Moser and Ekstrom, 2010) . Policies created for mitigation and adaptation in climate science for rural farmers do not go hand in hand with the farmers' attitude towards climate change. Besides, in Kenya and other East-African countries, notwithstanding the increase in climate information from scientific institutions on climate, there is no knowledge transfer of that data to decision makers (Bryan, Ringler, Okoba, Koo, et al., 2013) as well as other barriers. (Kabisch et al., 2016) posit the literature varies on determining barriers to mitigation and adaptation measures uptake. Drawing from the argument that mitigation and adaptation are different and can be distinguished by behavioral responses. Indeed, mitigation can be defined as a local response to an international need such as reduction of emission of Greenhouse gases, adaptation focuses on a local need and its response thereof (Bryan, Ringler, Okoba, Koo, et al., 2013) an example is food security. In this study, barriers are defined as that which stands in between climate problems as they are identified and their solution which is classified as climate science, and that generally concern itself around the three dimensions of social, biophysical and economic. Behavioral and institutional barriers make up the category of social barriers. The way institutional organizations and their interactions with individuals influence the way individuals are permitted to make changes up to some limit is the institutional barrier related category. Behavioral barriers, on the other hand, are specifically related to how thought processes influence the way individual actors react to climate change stimuli (Jones and Boyd 2011). Environmentally, environmental citizenship behavior is a vital factor in determining uptake of climate science which is Adaptation and mitigation. This behavior preserves and enhances the environment which in turn contributes to sustainable development. Therefore, factors influencing environmental citizenship behavior in this paper aim at promoting sustainable agriculture and consequently, sustainable development. According to the value-belief-norm (VBN) theory (P. C. Stern, Dietz, Abel, Guagnano, & Kalof, 1999), it is assumed that a consumers' value orientation (VO) affects their Ascribed Responsibility (AR). AR if moderated by Environmental Knowledge (EK), will be eventually expressed in environmental positive significance which is environmental citizenship behavior. In the Kenyan context, there is a limitation in research that look into the moderated relationship between AR and ECB. Moreover, activated knowledge (Priadi, Fatria, Sarkawi, & Oktaviani, 2018) has significance to stimulate environmentally significant behavior. Some scholars have drawn Schwartz' norm-activation theory of altruistic behavior for example Stern & Dietz, 1994; Stern, Dietz, & Black, 1986; Stern, Dietz, & (P. C. Stern et al., 1999); (Liere & Dunlap, 1981) directly apply to positive environmental behavior since these are inner morals and personally upheld norms . In the applications of Schwartz' theory, awareness of consequences catalyzes pro environmental behavior. This can be merely compared to environmental concern or positive attitude (Paul C Stern, 2000). In line with Berkman (2002) he distinguishes between awareness of ego oriented, social concern, and environmental consequences corresponding to three different underlying value orientations namely; biospheric, egoistic and altruistic. In order for the behavior to be performed, attention to or awareness of consequences must induce an Ascription to Responsibility (AR) to perform the behavior that in turn activates a personal norm or moral responsibility to perform the behavior.

The mediating effect of Ascribed responsibility between value orientation and ECB has not been looked into in the Sub-Saharan context. Additionally, the moderator used in my study has not been explored. The moderating effect shows human duty in providing environmental knowledge drives environmentally significant behavior (Liobikiene & Poškus, 2019).

### 1.1. Problem Statement

Common among countries in Sub-Saharan Africa , Kenya continuously faces the challenges of climate change (Recha, 2019). The country depends on rainfall agriculture, modern technology adoption has not been prioritized, poverty reduction had not been achieved, and infrastructure is poor thus markets least developed (Bryan, Ringler, Okoba, Koo, et al., 2013) (Odhiambo, 2009) .Climate models suggest that the Kenyan Average temperature is higher than most regions worldwide. However, uncertainty looms wide about future changes in rainfall in the region. While many universal models show minimal rise in average precipitation in East Africa (Bryan, Ringler, Okoba, Roncoli, et al., 2013) these may be offset by warming of the Indian ocean, more rainfall variations and more occurrences of extreme events such as droughts. (Eisenack et al., 2014) Regional variations in precipitation are more ; It is assumed and expected to get wetter in the Kenyan highlands and Northern Kenya which differs from the coastal region and lowlands which usually get drier(Bryan, Ringler, Okoba, Roncoli, et al., 2013) . Furthermore, key staples like maize and wheat may reduce despite there being more rain owing to increased evapotranspiration (Tidwell, 2010).

This paper researches the relationship between agricultural productivity, GHG mitigation in Kenya based on farming practices being used by farmers. The main moderating factor introduced in this paper is Environmental knowledge. Policymakers can be influenced by the results to implement policies that encourage better management practices which are effective as well as available in achieving different management practices for the various agroecological zones (AEZs) in Kenya and beyond.

### 1.2. Local Ecological Knowledge

Smallholder farmers are vital in agricultural improvement (Tidwell, 2010 & FAO, 2008). In targeting these farmers and strategies for adaptation on their farm management practices or Climate smart Agriculture, their local knowledge is key to up their adaptive capacity as seen in the literature for climate change, traditional ecological knowledge (Pretty et al., 2009) . Both local knowledge and local ecological knowledge refer to the a collection of knowledge, practices and beliefs, that are within a particular locality, only reached through a long-term observation of while being present in that environment, and transferred through oral traditions through generations (Ogalleh, Vogl, Eitzinger, & Hauser, 2012).

Many smallholders often apply this local ecological knowledge in their daily practices. In assessing adaptive capacity, it is vital to look at these practices to improve local knowledge. This illustrates how to merge agricultural practices to up smallholders' adaptive capacity towards climatic variations within the seasons. This knowledge is specific to location, time and culture(Bank & Bank, 2019). Local knowledge is practical and farmers decide wisely and while more informed at certain times to environmental changes and how to make their yield better(Saitabau & Nairobi-kenya, 2014). The test of validity of local knowledge has been proven by science by comparing with quantitative data (Orlove & Caton, 2010). Many results illustrates that farmers' observations matches quantitative data analysis: local knowledge has been used to respond to extreme conditions which farmers have encountered such as droughts, famines, and other condition (Jiri, Mafongoya, Mubaya, & Mafongoya, 2016). There have also been cases where there has been a mismatch between local knowledge and quantified data hence casting doubt on local knowledge (Ochieng, Recha, & Bebe, 2017)

Policy on adaptation policy has not been grounded yet. Despite all the noise on climate change policy agenda, only used during political campaigns, adaptation policy research is not adapted enough .Focus is on how to measure and scale-up to the "dependent variable problem"(Dupuis & Biesbroek, 2013). While the literature on adaptation has dealt with the "how" to understand adaptation concepts from different angles such as resilience, adaptive capacity and vulnerability (Lee, Yang, & Blok, 2020), questions loom on the practicality of readiness for future climate change. Research uptake thus still remains an emerging field that is not yet very well understood.

Collaborations such as "boundary organizations" try to make scientists and policy-makers have exchange forums for exchange of information, policy learning and decision-making processes (Lee et al., 2020)

The funding for climate modelling and the quality and quantity of climate data available is very variable on a global scale and in Kenya.

## II. THEORETICAL BACKGROUND AND MODEL

### 2.1. Value Belief Norm (VBN) Theory

Based on the value-belief-norm (VBN) theory (Paul C Stern, 2000), For a consumer to have environmental significant behavior, he must have values which affect his beliefs that guide his pro-environmental personal Norm (PPN). PPN obliges one to act pro environmentally (Sponarski, Vaske, & Bath, 2015; Wolf, 1958). Belief is the acceptance that nature is factual and accurate and should be kept as so. According to some authors, values are defined to be concepts acceptable behavior that go beyond normalcy  (Bruvold, 1973 & Chua et al., 2019) . (Schwartz, 1992) aligns value orientation to what is valuable to humans. Stern et al. (1998) adopted a socio psychological perspective in defining the value orientations. Stern (2000) later added the VO, belief and norm on environmentally significant behavior to create his VBN theory. This study purposes to adopt the said values from the VBN theory and enhance the same using Environmental knowledge to measure the impact of the Environmental knowledge of farmers on the uptake and use of scientific research methods on mitigation and adaptation.

### 2.2. Environmental Citizenship Behavior (ECB) Model

Some observers define the model predominantly as a "pro-environmental behavior. Andrew Dobson (2010), is of the view that the model argues the principle of fairness and the sharing of what the environment provides, by taking part in and in the co-creation of laws that sustain development where the scope of citizen taking part ranges from individuals personally taking part in environmental decision making process or by using influence to directly sort out environmental concerns and to work with others to reach solutions  for environmental crises.

In Agriculture, ECB guides to comprehend environmentally significant behavior(Boon, Quoquab, & Mohammad, 2019).

Environmental citizenship encompasses both activist and non-activist support for support good environment behavior (Paul C Stern, 1995). Individuals may not directly be seen to be promoting pro-environmental behavior but may support groups or even join such initiatives(Abedinpour et al., 2012)in non-activism support. (Van Herzele et al., 2013) observes that to understand environmentally significant behavior in Agriculture, ECB is a key component.  Farmers are businessmen who need to be efficient (Del Corso, Kephaliacos, & Plumecocq, 2015) They use all resources available to spread awareness of good environmental behavior. As such, pro-environmentalists preserve and improve the condition of their environment (Gailhard, Bavorová, & Pirscher, 2015) They also form group with subscription membership. Farmers have a tendency to pressure their elected leaders to protect their interests (Taylor & Van Grieken, 2015) In Kenya, they are seen to constantly pursue the government for subsidized fertilizer prices and irrigation initiatives.

ECB is considered as a collective effort which may also include attending seminars(Paul C Stern, 2000).

### 2.3. Value orientation

Schwartz (1992) says that values improve existent status or are acceptable. (P. C. Stern et al., 1999) adopted measures for value orientation in a socio psychological perspective from specific values in society to study good environmental behavior.

Stern (2000) later added the Value Orientation to and belief and norm to create the VBN theory on environmentally significant behavior. BV Is the farmer's willingness to protect his environment or biosphere without harm (Turaga, Howarth, & Borsuk, 2010). It unites man and his natural resources (Steg, 2007) and non-pollution due to respect of the environment (Berkman, 2002). Emphasis is put on preservation of the environment (Steg, 2007) . Farmers should highly perform Biospheric values (Del Corso et al., 2015).

The Altruistic Value is more inclined towards the farmer taking care of the environment so as not to harm the society and other people (Chen & Sun, 2015a) (Chua et al., 2019). Farmers who have a high altruistic value have environmentally significant behavior (Del Corso et al., 2015).

EV are those for the personal benefit of an individual (Turaga et al., 2010) . the individual considers his needs before safeguarding his environment (Chen & Sun, 2015b)  are the basis of EV.

Relationship between value orientation and Ascribed responsibility

AR is believed to be human action impact on the environment. It could be positive or negative (Paul C Stern, 2000). The VBN gives an understanding of the relationship between VO and AR. The relationship between humans and their environment may have consequences depending on how humans treat their environment. (Saleem, Eagle, & Low, 2018)(Fielding, McDonald, & Louis, 2008)(Snelgar, 2006) In this case, farmers are aware that if they use certain agrochemicals, it will be harmful to their environment and as such will desist from using them

H1. Biospheric value has a direct positive effect on Ascribed responsibility

H2.  Altruistic Value has a direct positive effect on Ascribed responsibility

H3. Egoistic value has a direct negative effect on Ascribed responsibility

H4. Ascribed responsibility has a direct positive effect on Environmental Citizenship Behavior

Environmental knowledge may lead to pro-environmental behavior especially if there is monitoring of such knowledge. A few scholars have found that such knowledge has an insignificant effect on pro-environmental behavior . (Otto & Kaiser, 2014) However found this relationship to be significant. Their study found that people who know more about environmental problems and have vast knowledge on the environment act more pro-environmentally. This inconsistency is proof that there has to be studies on more than one school of thought on the impact of Environmental knowledge  (Liobikiene & Poškus, 2019),

The effect of Action-related knowledge has a greater impact on pro-environmental compared to any other type of environmental knowledge. For example, merely knowing that climate smart agriculture has a better effect on the environment is not enough but knowing that the farmer will get better yields can influence policy decisions. This affects behavior directly (Liobikiene & Poškus, 2019). Goes without saying that people with higher knowledge behave more appropriately towards the environment as stated by Ting (Otto & Kaiser, 2014). (Zhao, Gao, Wu, Wang, & Zhu, 2014) observed that usage behavior can only be achieved if and when one is informed about green economy. The knowledge of environmental problems impacts pro-environmental behavior(López & Cuervo-arango, 2014). One can only take pro-environmental actions if they know what the can or cannot do. Thus, action related knowledge translates to behavior.

### 2.4. The relationship between Ascribed Responsibility and environmental citizenship behavior with the moderating role of Environmental knowledge.

Independent and Dependent variables can be perfectly described where there is a moderator. The current study proposes EK to moderate the relationship between AR and ECB. (Steg, 2007) that personal norms positive to the environment mediated the relationship between environmentally significant behavior and the ascription of responsibility. The authors did an experiment on the pricing policy for transport and its acceptance. The nexus between the ascription of responsibility, environmental knowledge (EK) and ECB is yet to be examined. The current study assumes that EK will moderate the effect of AR on ECB positively. For this reason, the hypothesis hereunder is proposed.

H5: Environmental Knowledge moderates Ascribed Responsibility positively

From the foregoing, the figurative model below is proposed.

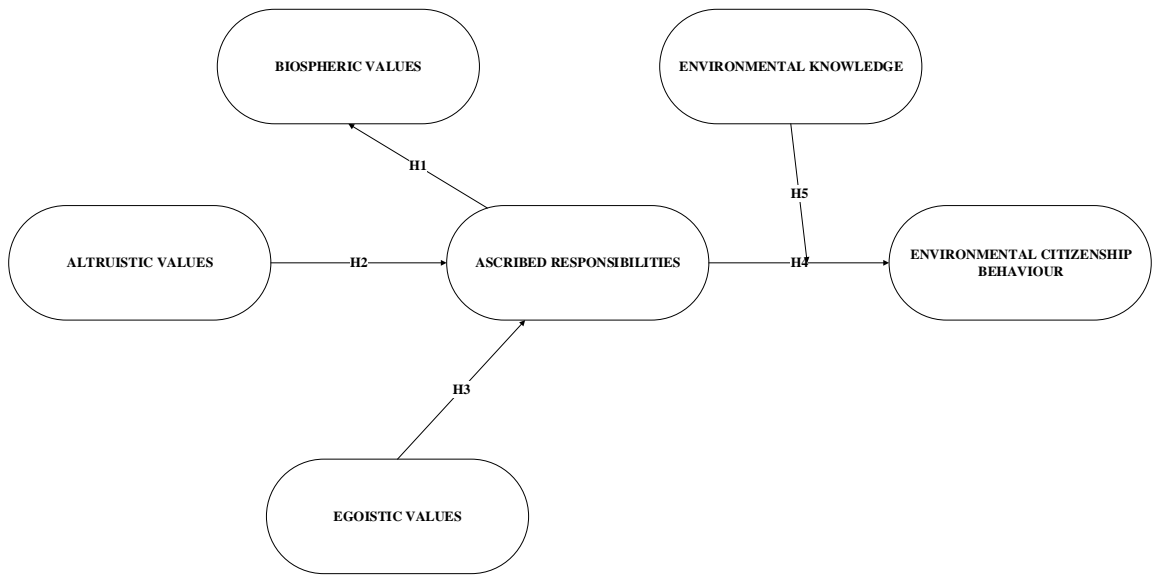

*Fig.1: Conceptual Model*

## III. METHODOLOGY
### 3.1. The Study Area

Kenya's total area of 581,700 km2consists ranges in natural ecosystems. There are the following areas; arid, semi-arid, savannah and forests. There is rapid and unending expansion of urban areas despite urbanization and of rural areas. Areas covered by water and the ocean environment which hosts marine and coastal ecosystems, fresh water lakes and rivers. Some of these rivers are periodic and may dry at different periods during climate variability. Some of the lakes are saline in nature while some are of fresh water. The aquatic environment includes 14,300 and 143,100 km2 of territorial waters and exclusive economic zone (EEZ), respectively, in the Indian Ocean (Services, Health, & Macro, 2010) . Table 1 shows the different Agro ecological zones of the country. Kenya's economy is dependent on the environment. All the key sectors, rely on the environment. To achieve a sustainable economy and development, different sectors of development must take action. Agriculture has for a long time been the main contribution to the economy marking up for 25 percent of the gross domestic product (GDP) (Kabubo-Mariara & Karanja, 2007) .

*Table 1 Agro- ecological zones of Kenya*

| Zone | Approximate Area (km2) | % Total |
|---|---|---|

| I. Agro-Alphine | 800 | 0.1 |
| --- | --- | --- |
| II. High Potential | 53,000 | 9.3 |
| III. Medium Potential | 53,000 | 9.3 |
| IV. Semi-Arid | 48,200 | 8.5 |
| V. Arid | 300,000 | 52.9 |
| VI. Very arid | 112,000 | 19.8 |
| Rest (waters etc) | 15,600 | 2.6 |

Source: Sombroek, et al., 1982.

### 3.2. Statistical Analysis

A scholar or researcher may use mediation analysis to realize the effect of X on Y (Hayes & Rockwood, 2017). The authors here use the example of the therapy and trauma. They explain that the initial cause X could be the kind of the therapy the patient receives or the kind of therapy or any other conceived actor that has some kind of cause which consequently has results.

Whilst mediation analysis focuses the cause and its consequences, moderation analysis deals with, time and conditions, or for the kind of individuals that effect exists or does not and to what extent.(Hayes & Rockwood, 2017). In this paper, the authors used the example of traditional therapy in comparison to a modern therapy which effects might be less effective to treating depression. The effect of the moderation might be small or even harmful.

In this paper Environmental knowledge will be the moderator on ascribed responsibility

In the collection of data, the research identified and assesses current and potential household-level Adaptation and Mitigation strategies available to farmers from 4 different Agro ecological zones (AEZs) was collected, cash crops, other crops, institutional backgrounds and policy were all considered. World Bank supported projects in the selected zones where agricultural mitigation and adaptation were practiced. They range from arid, semi-arid, temperate, and humid areas.

### 3.3. Data collection

Participants

The use of self-administered questionnaires was employed by distributing 350 questionnaires to those farmers who had basic knowledge on Climate smart agriculture. Around 300 were completed usably and handed back directly by the farmers after completion.

*Table 2 study sites*

| district | Agro ecological zone | No of Households |
| --- | --- | --- |
| Garissa | Arid | 75 |
| Njoro | Semi-arid | 75 |
| Othaya | Temperate | 100 |
| Siaya | Humid | 100 |
| **Total** |  | **350** |

### 3.4. Study Design

The hypothetic-deductive approach was followed in this study, wherein hypotheses were tested. Each variable correlation is tested without much engagement of the researcher. Positive Environmental behavior was observed as it usually and normally occurs. The farmer being the unit of analysis. The data were collected in the year 2019 which has 2 main planting seasons however some plants especially vegetables have no particular season thus some grow throughout the year. The objective of this study being to confirm existing theory (Value Belief Norm) by developing new variables in this case, moderating Ascribed responsibility using Environmental knowledge; non-probability sampling technique, was applied (Boon et al., 2019). The thumb rule was used to reach the appropriate sample size (Cracraft, 1988).

### IV. RESULTS AND ANALYSIS

A five-point Likert scale ranging from strongly disagree (never behave: 1) to strongly agree (always behave: 5) is employed. Six variables were investigated in this study: Biospheric value, Altruistic value, Egoistic value, environmental knowledge, Ascribed responsibility and Environmental Citizenship Behavior. The scale for the 3 value orientations were measured according to stern in (P. C. Stern et al., 1999). They were measured using 9 items such as; While farmers use the environment, the onus lies in them protecting it (e.g. "I only use environmentally friendly fertilizer", "I long for a war free world" and "I would like to have better harvest next time for more monetary returns"). To measure environmental knowledge ,reference was made to (Frick, Kaiser, & Wilson, 2004) who focused more on action-related knowledge to reveal the real knowledge about the impact of a specific action. Scales for Ascribed responsibility was constructed by adopting items used by (Paul C Stern, 1995) . "disposal of agricultural waste has contributed to increase of cancer deaths " and "No one has the right to harm the environment" . Environmental Citizenship Behavior was measured by adapting the items suggested by (Paul C Stern & Dietz, 1994) items measured were 3 ("I always watch out my elected leader's contribution regarding environment issues related to irrigation methods and water conservation methods.)

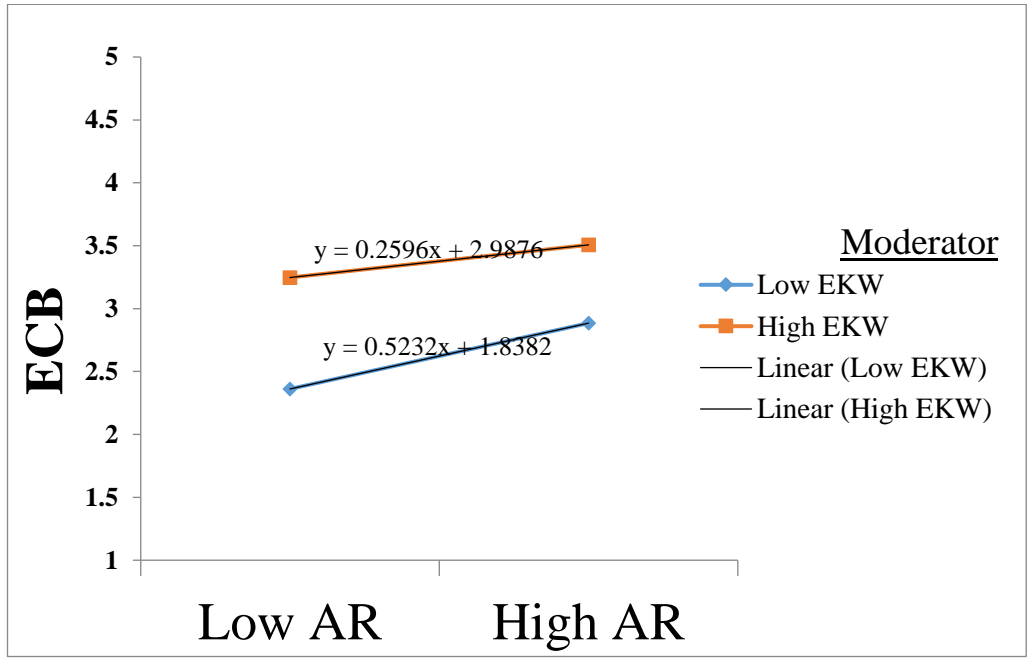

*Fig 2: EKW dampens the positive relationship between AR and ECB*

*Table 3 ANOVA and Table 4. Regression Coefficient*

| ANOVA[a] | | | | | | |
|---|---|---|---|---|---|---|
| Model | | Sum of Squares | df | Mean Square | F | Sig. |
| 1 | Regression | 56.285 | 1 | 56.285 | 176.183 | .000[b] |
| | Residual | 104.787 | 328 | .319 | | |
| | Total | 161.072 | 329 | | | |
| 2 | Regression | 66.413 | 2 | 33.206 | 114.711 | .000[c] |
| | Residual | 94.659 | 327 | .289 | | |
| | Total | 161.072 | 329 | | | |

a. Dependent Variable: ECB

b. Predictors: (Constant), EKW

c. Predictors: (Constant), EKW, AR

**Model Results**

| Model | R | R Square | Adjusted R Square | Std. Error of the Estimate | Change Statistics R Square Change | F Change | df1 | df2 | Sig. F Change |
|---|---|---|---|---|---|---|---|---|---|
| 1 | .591[a] | .349 | .347 | .56522 | .349 | 176.183 | 1 | 328 | .000 |
| 2 | .642[b] | .412 | .409 | .53803 | .063 | 34.985 | 1 | 327 | .000 |

a. Predictors: (Constant), EKW

b. Predictors: (Constant), EKW, AR

Table 4 model of the regression coefficient shows F (1, 328) = 176.183 p < = 0.001 and F (2, 327) = 34. 985 p < = 0.001. per this regression output shows potential significance. Table 3 ANOVA also further indicates that, the interaction of EKW accounted for the significant variance of AR and ECB with R² = 0.063 significance potential moderation between AR and ECB. The moderation interaction from Figure 2 also indicates dampen positive relationship between AR and ECB. Both model indicates significance of the variance, signifies moderation perceived in the hypothesis 5.

*Table 4. Hayes' Process condition output of mediation.     \*p < = .01, \*\*p < = .001, \*\*\*p < = .005. .*

| OUTCOME VARIABLE: | of MEDIATION IN THE STUDY | | | | | |
|---|---|---|---|---|---|---|

| Model Summary | Exogenous construct predicts | | significance predicts Mediation AR | | | |
|---|---|---|---|---|---|---|
| R | R-sq | MSE | F | df1 | df2 | p |
| .3627 | .1315 | .6759 | 16.4588 | 3.0000 | 326.0000 | .0000 |

|  | coeff | se | t | p | LLCI | ULCI |
|---|---|---|---|---|---|---|
| constant | 2.1904 | .3947 | 5.5494 | .0000 | 1.4139 | 2.9669 |
| EV | -.1260 | .0592 | -2.1265 | .0342 | -.2425 | -.0094 |
| AV | .2207 | .0552 | 4.0012 | .0001 | .1122 | .3292 |
| BV | .3251 | .0670 | 4.8523 | .0000 | .1933 | .4569 |

| Model Mediation/exogenous significance predictor of endogenous construct | | | | | | |
|---|---|---|---|---|---|---|
| R | R-sq | MSE | F | df1 | df2 | p |
| .5981 | .3577 | .3183 | 45.2440 | 4.0000 | 325.0000 | .0000 |

**Model**

|  | coeff | se | t | p | LLCI | ULCI |
|---|---|---|---|---|---|---|
| constant | 1.4483 | .2834 | 5.1106 | .0000 | .8908 | 2.0058 |
| EV | -.0870 | .0409 | -2.1252 | .0343 | -.1675 | -.0065 |
| AR | .2548 | .0380 | 6.7050 | .0000 | .1801 | .3296 |
| AV | .1960 | .0388 | 5.0553 | .0000 | .1197 | .2723 |
| BV | .2973 | .0476 | 6.2438 | .0000 | .2036 | .3910 |

Test(s) of X by M interaction:

| F | df1 | df2 | p |
|---|---|---|---|
| 7.5082 | 1.0000 | 324.0000 | .0065 |

Indirect effect(s) of X on Y:     exogenous significant predicts endogenous constructs

| | Effect | BootSE | BootLLCI | BootULCI |
|---|---|---|---|---|
| AR | -.0321 | .0169 | -.0682 | -.0020 |

Partially standardized indirect effect(s) of X on Y:

| | Effect | BootSE | BootLLCI | BootULCI |
|---|---|---|---|---|
| AR | -.0459 | .0232 | -.0940 | -.0028 |

Completely standardized indirect effect(s) of X on Y:

| | Effect | BootSE | BootLLCI | BootULCI |
|---|---|---|---|---|
| AR | -.0361 | .0182 | -.0741 | -.0023 |

Table 4 indicates mediation of exogenous variables of (EV, AV and BV) to AR with indirect effect on ECB significant of ($\beta$ = .0592, t = -2.1265, p < = .0342), ($\beta$ = .0552, t = 4.0012, p < = .001), and ($\beta$ = .0670, t = 4.8523, p < = .000) respectively. The regression model shows all exogenous variables significance predictor of AR except EV. In the same vein, the indirect effect of (EV, AV, BV) significantly predicts the ECB shows table 4 EV ($\beta$ = .0409, t = -2.1252, p < = .0343), AR ($\beta$ = .0380, t = 6.7050, p < = .000), AV ($\beta$ = .0388, t = 5.0553, p < = .000), BV ($\beta$ = ..0476, t = 6.2438, p < = .000). The effect size is -.0361 with 95% confidence level thus less than zero in negative interval z = -.002(Preacher & Hayes, 2008) . Therefore, the H4 of the indirect effects of all the exogenous constructs significantly satisfied the model constructs in this study. Table 5 shows very good significance relationship among constructs, though, on negative weak relationship but significant. Therefore, all the constructs have showed good effects from H1, H2, and H3 .(Chua et al., 2019)

The hypotheses in the study all supported, while the mediation and moderation also justified from table 4 and 3 above. From table 6 items loading from each construct indicated above or within the benchmark of 0.7. These loading are indication of good measurement effects of the exogenous variables and the endogenous constructs based on the conceptual model of the study. The variance explanatory power of ECB is $R^2$ = 0.650 represents 66% of the dependent variable strength. Also, the mediator $R^2$= 0.363 thus 36% of the variance of all the exogenous directly to the mediator.

*Table 5: Means, standard deviations (SD) and Pearson correlations (r)*

| Variables | EV | AV | BV | ECB | AR | EKW | @ | Mean | SD |
|---|---|---|---|---|---|---|---|---|---|
| **EV** | 1 | | | | | | 0.706 | 3.5733 | 0.7859 |
| **AV** | -0.166** | 1 | | | | | 0.706 | 3.9356 | 0.83896 |
| **BV** | 0.140* | 0.09 | 1 | | | | 0.891 | 4.1642 | 0.68788 |
| **ECB** | -0.132* | 0.358** | 0.382** | 1 | | | 0.905 | 4.1508 | 0.6997 |
| **AR** | -o.112* | 0.253** | 0.258** | 0.466** | 1 | | 0.857 | 3.9629 | 0.87819 |
| **EKW** | -o.174** | 0.411** | .327** | 0.591** | 0.398** | | 0.891 | 4.1568 | 0.79189 |

Correlation is significant at the * = P≤0.05, ** = P≤0.01, Note: @ = Cronbach's Alpha, SD = standard deviation, EGV = egoistic values, ALV = altruistic values, BPV = biospheric values, AR = ascribe responsibilities, ECB = environmental citizenship behavior, EKW = environmental knowledge.

*Table 6 of factor loading*

**Loading of the constructs on the measured items**

| ITEMS | 1 | 2 | 3 | 4 | 5 | 6 |
|---|---|---|---|---|---|---|
| AV3 | **0.866** | 0.088 | 0.154 | 0.123 | 0.015 | -0.019 |
| AV2 | **0.860** | 0.050 | 0.212 | 0.092 | -0.003 | -0.083 |
| AV4 | **0.820** | 0.127 | 0.084 | 0.175 | 0.048 | -0.065 |
| AV1 | **0.800** | 0.090 | 0.110 | 0.097 | 0.004 | -0.094 |
| AR3 | 0.064 | **0.881** | 0.164 | 0.145 | 0.072 | 0.003 |
| AR1 | 0.048 | **0.816** | 0.158 | 0.172 | 0.118 | -0.082 |
| AR2 | 0.144 | **0.814** | 0.050 | 0.195 | 0.106 | -0.093 |
| AR4 | 0.111 | **0.809** | 0.152 | 0.137 | 0.080 | 0.010 |
| EKW3 | 0.160 | 0.136 | **0.861** | 0.235 | 0.122 | -0.052 |
| EKW4 | 0.101 | 0.211 | **0.821** | 0.144 | 0.123 | -0.120 |
| EKW2 | 0.218 | 0.089 | **0.775** | 0.320 | 0.105 | -0.091 |
| EKW1 | 0.236 | 0.194 | **0.707** | 0.285 | 0.192 | -0.041 |
| ECB3 | 0.152 | 0.192 | 0.238 | **0.812** | 0.147 | -0.009 |
| ECB2 | 0.188 | 0.180 | 0.287 | **0.787** | 0.131 | -0.048 |
| ECB1 | 0.205 | 0.214 | 0.178 | **0.783** | 0.195 | 0.012 |
| ECB4 | 0.055 | 0.210 | 0.259 | **0.715** | 0.220 | -0.168 |
| BV2 | 0.030 | 0.036 | 0.053 | 0.163 | **0.831** | 0.082 |
| BV4 | 0.011 | 0.083 | 0.154 | 0.093 | **0.815** | 0.067 |
| BV3 | 0.076 | 0.175 | 0.143 | 0.086 | **0.813** | 0.140 |
| BV1 | -0.049 | 0.078 | 0.074 | 0.188 | **0.777** | -0.030 |
| EV2 | -0.137 | -0.003 | -0.038 | -0.010 | 0.029 | **0.834** |

| EV4 | -0.020 | -0.014 | -0.148 | -0.032 | 0.072 | **0.754** |
| EV1 | -0.231 | -0.076 | -0.179 | 0.038 | -0.026 | **0.748** |
| EV3 | 0.155 | -0.067 | 0.168 | -0.189 | 0.203 | **0.538** |

Extraction Method: Principal Component Analysis.
Rotation Method: Varimax with Kaiser Normalization.

a. Rotation converged in 6 iterations.

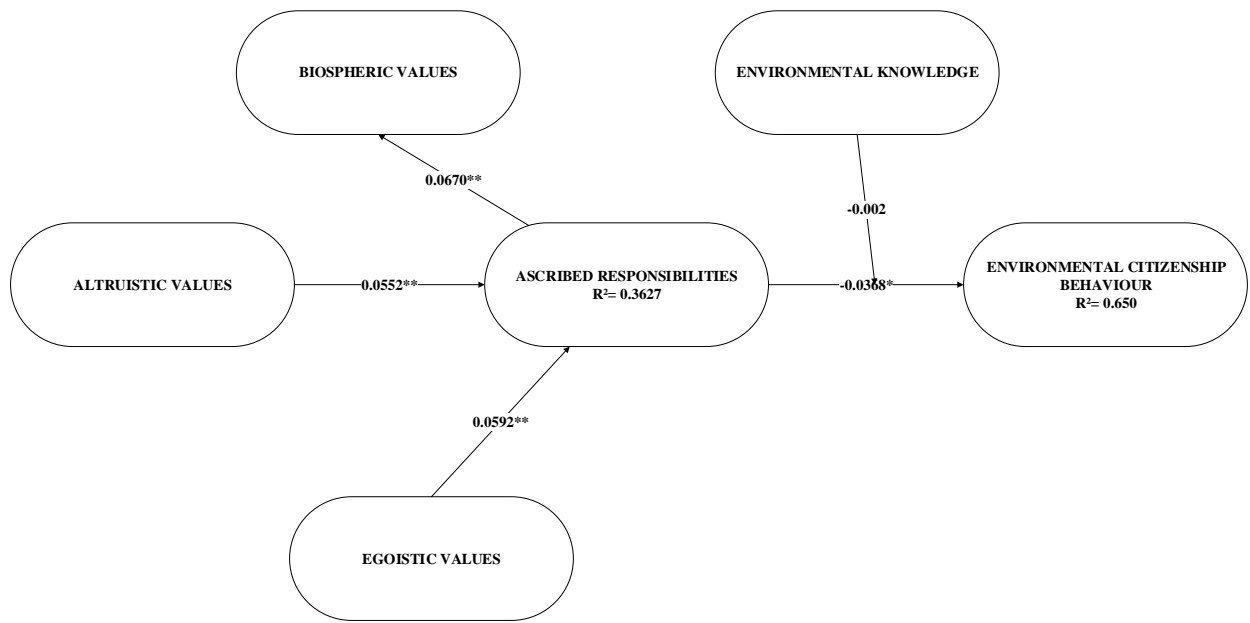

*Fig.3: Regression path coefficient of the conceptual model*

## V. DISCUSSIONS

The current study looks into the nexus between Value Orientation and Environmental Citizenship Behavior while testing the mediating effect of Ascribed Responsibility and Moderating effect of Environmental Knowledge. On the ground of VBN theory, hypotheses were developed and thereafter tested using the advanced statistical technique, i.e., PLS-SEM. The results confirmed all hypotheses context of Kenyan farmers. It was important to examine different Agro ecological zones because of the varying temperatures and soil humidity. The synergy between value orientation and Ascribed responsibility in this study, VO in terms of BV and AV positively and significantly affected AR. EV on the other hand yields negative results. VBN theory therefore confirmed, which argued hypothesis of values are formed to prove a direct effect on how form and articulate AR (Paul C Stern, 2000), Thus, a farmer with strong BV and AV is expected to accept AR and speedily welcome initiatives. The farmers consider their environment while using fertilizer, they are concerned about the wellbeing of their fellow citizens and they focus on their harvest. When the farmers are egoistic though, they don't seem to care much about the impacts of their action. These value orientations affect Ascribed responsibility, and show how much they can do for environmental protection.

### 4.1. The relationship between Ascribed Responsibility and Environmental citizenship Behavior

Data obtained confirms the connection between AR and the ECB as is the case in the VBN theory. Individuals aware of what their actions might cause and capable of exercising caution refrain from harmful actions to the environment. The current research context, Kenyan farmers are more knowledgeable and employ the use of using Climate Smart methods in farming and consequently intimated the will to carry out their responsibility to take care of the environment. Therefore, Responsibility is confirmed. On Environmental knowledge, the current study revealed that EK enhances a farmer's responsibility to be more aware and as such have the final effect in decision making. The results are true to the VBN theory, which assumes a nexus between variables and action taken, i.e. values, responsibility, norm and positive environmental behavior (Paul C Stern, 2000)Farmers who are aware of action related consequences and know their commitment to the environment, have their sense of moral obligation activated, and eventually influences their Environmental Citizenship Behavior.

## VI. CONCLUSIONS AND POLICY IMPLICATIONS

This study analyzes behavior of farmers towards adaptation and mitigation regarding climate change showing their knowledge of effects of climate change and show that some farmers have employed various adaptation methods. There is also a difference in effects in the different Agro Ecological Zones. Climate Smart Agriculture such as Crop rotation is mostly used in high potential zones, while irrigation and water harvesting are more common in dryer regions. If Kenyan farmers intend to counter the hard long term effects of climate change, more environmental knowledge on climate change must be instilled.

Provision of Environmental knowledge to farmers by the government should be a key policy issue. Monitoring and evaluation of practices on the knowledge to farmers should also be a key factor for adaptation to climate change. The gap in different disciplines such as scientists, climate experts and policy makers should be filled to disseminate knowledge. Scientists may spend all the time in the lab and have results but if not implemented into policy then it's a waste of government resources especially if the scientists are government funded.

Access to credit by farmers may be used as a key tool to attain adaptation measures. The same should not attract high interest rates so as not to discourage farmers.

Action-based knowledge does not predict public sphere behavior but Public behavior pretty much determines private sphere behavior. Suffice to say, if behavior is determined by public law, it streams down to private individual behavior. This means that Environmental citizenship behavior impacts values which may be changed because of awareness. Knowledge and motivation encourage Environmental Citizenship Behavior and the same acts backwards. Therefore, environmental education should provide information as well as nature- based experiences to make it eco centric.

Livestock farming has not been considered in this paper yet most Kenyan farmers combine both crop and livestock farming. The arid and semi-arid farmers mainly concentrate on livestock rearing and this may be good for future research.

## ACKNOWLEDGEMENT

This is the original work of the authors who declare no conflict of interest nor funding associated.

## REFERENCES

[1] Abedinpour, M., Sarangi, A., Rajput, T. B. S., Singh, M., Pathak, H., & Ahmad, T. (2012). Performance evaluation of AquaCrop model for maize crop in a semi-arid environment. *Agricultural Water Management*, *110*, 55–66. https://doi.org/10.1016/j.agwat.2012.04.001

[2] Ayeri, O. S., Christian, V. R., Josef, E., & Michael, H. (2012). Local perceptions and responses to climate change and variability: The case of Laikipia District, Kenya. *Sustainability*, *4*(12), 3302–3325. https://doi.org/10.3390/su4123302

[3] Bank, M. Van Der, & Bank, C. M. Van Der. (2019). *A Legal Opinion on Mitigation and Adaptation on Air Pollution Strategies for Local Governments in South Africa*. *13*(10), 1306–1311.

[4] Berkman, P. A. (2002). Ecosystem Conservation. *Science Into Policy*, *279*(April), 157–181. https://doi.org/10.1016/b978-012091560-6/50013-5

[5] Boon, C. K., Quoquab, F., & Mohammad, J. (2019). *Factors affecting environmental citizenship behaviour An empirical investigation in*. (July). https://doi.org/10.1108/APJML-05-2018-0177

[6] Bryan, E., Ringler, C., Okoba, B., Koo, J., Herrero, M., & Silvestri, S. (2013). *Can agriculture support climate change adaptation , greenhouse gas mitigation and rural livelihoods ? insights from Kenya*. *00800*, 151–165. https://doi.org/10.1007/s10584-012-0640-0

[7] Bryan, E., Ringler, C., Okoba, B., Roncoli, C., Silvestri, S., & Herrero, M. (2013). Adapting agriculture to climate change in Kenya: Household strategies and determinants. *Journal of Environmental Management*, *114*, 26–35. https://doi.org/10.1016/j.jenvman.2012.10.036

[8] Chen, H., & Sun, J. (2015a). *Changes in climate extreme events in China associated*. *2751*(September 2014), 2735–2751. https://doi.org/10.1002/joc.4168

[9] Chen, H., & Sun, J. (2015b). Changes in climate extreme events in China associated with warming. *International Journal of Climatology*, *35*(10), 2735–2751. https://doi.org/10.1002/joc.4168

[10] Chua, K. B., Quoquab, F., & Mohammad, J. (2019). Factors affecting environmental citizenship behaviour: An empirical investigation in Malaysian paddy industry. *Asia Pacific Journal of Marketing and Logistics*, *32*(1), 86–104. https://doi.org/10.1108/APJML-05-2018-0177

[11] Cracraft, J. (1988). Early evolution of birds. *Nature*, *331*(6155),

389–390. https://doi.org/10.1038/331389a0
[12] Del Corso, J. P., Kephaliacos, C., & Plumecocq, G. (2015). Legitimizing farmers' new knowledge, learning and practices through communicative action: Application of an agro-environmental policy. *Ecological Economics*, *117*, 86–96. https://doi.org/10.1016/j.ecolecon.2015.05.017
[13] Dupuis, J., & Biesbroek, R. (2013). Comparing apples and oranges: The dependent variable problem in comparing and evaluating climate change adaptation policies. *Global Environmental Change*, *23*(6), 1476–1487. https://doi.org/10.1016/j.gloenvcha.2013.07.022
[14] Eisenack, K., Moser, S. C., Hoffmann, E., Klein, R. J. T., Oberlack, C., Pechan, A., … Termeer, C. J. A. M. (2014). Explaining and overcoming barriers to climate change adaptation. *Nature Climate Change*, *4*(10), 867–872. https://doi.org/10.1038/nclimate2350
[15] Fielding, K. S., McDonald, R., & Louis, W. R. (2008). Theory of planned behaviour, identity and intentions to engage in environmental activism. *Journal of Environmental Psychology*, *28*(4), 318–326. https://doi.org/10.1016/j.jenvp.2008.03.003
[16] Frick, J., Kaiser, F. G., & Wilson, M. (2004). Environmental knowledge and conservation behavior: Exploring prevalence and structure in a representative sample. *Personality and Individual Differences*, *37*(8), 1597–1613. https://doi.org/10.1016/j.paid.2004.02.015
[17] Gailhard, İ. U., Bavorová, M., & Pirscher, F. (2015). Adoption of agri-environmental measures by organic farmers: The role of interpersonal communication. *Journal of Agricultural Education and Extension*, *21*(2), 127–148. https://doi.org/10.1080/1389224X.2014.913985
[18] Hayes, A. F., & Rockwood, N. J. (2017). Regression-based statistical mediation and moderation analysis in clinical research: Observations, recommendations, and implementation. *Behaviour Research and Therapy*, *98*, 39–57. https://doi.org/10.1016/j.brat.2016.11.001
[19] Jiri, O., Mafongoya, P. L., Mubaya, C., & Mafongoya, O. (2016). Seasonal Climate Prediction and Adaptation Using Indigenous Knowledge Systems in Agriculture Systems in Southern Africa: A Review. *Journal of Agricultural Science*, *8*(5), 156. https://doi.org/10.5539/jas.v8n5p156
[20] Kabisch, N., Frantzeskaki, N., Pauleit, S., Naumann, S., Davis, M., Artmann, M., … Bonn, A. (2016). Nature-based solutions to climate change mitigation and adaptation in urban areas: Perspectives on indicators, knowledge gaps, barriers, and opportunities for action. *Ecology and Society*, *21*(2). https://doi.org/10.5751/ES-08373-210239
[21] Kabubo-Mariara, J., & Karanja, F. K. (2007). The economic impact of climate change on Kenyan crop agriculture: A Ricardian approach. *Global and Planetary Change*, *57*(3–4), 319–330. https://doi.org/10.1016/j.gloplacha.2007.01.002
[22] Lee, T., Yang, H., & Blok, A. (2020). Does mitigation shape adaptation ? The urban climate mitigation-adaptation nexus. *Climate Policy*, *0*(0), 1–13. https://doi.org/10.1080/14693062.2020.1730152

[23] Liere, K. D. Van, & Dunlap, R. E. (1981). *Environment and Behavior*. https://doi.org/10.1177/0013916581136001
[24] Liobikiene, G., & Poškus, M. S. (2019). The importance of environmental knowledge for private and public sphere pro-environmental behavior: Modifying the Value-Belief-Norm theory. *Sustainability (Switzerland)*, *11*(12). https://doi.org/10.3390/su10023324
[25] López, A. G., & Cuervo-arango, M. A. (2014). *Relationship among values , beliefs , norms and ecological behaviour*. (May).
[26] Ochieng, R., Recha, C., & Bebe, B. O. (2017). *Enabling Conditions for Improved Use of Seasonal Climate Forecast in Arid and Semi-Arid Baringo County — Kenya*. *4*, 1–14. https://doi.org/10.4236/oalib.1103826
[27] Odhiambo, N. M. (2009). Energy consumption and economic growth nexus in Tanzania: An ARDL bounds testing approach. *Energy Policy*, *37*(2), 617–622. https://doi.org/10.1016/j.enpol.2008.09.077
[28] Ogalleh, S. A., Vogl, C. R., Eitzinger, J., & Hauser, M. (2012). *Local Perceptions and Responses to Climate Change and Variability: The Case of Laikipia District, Kenya*. 3302–3325. https://doi.org/10.3390/su4123302
[29] Orlove, B., & Caton, S. C. (2010). Water sustainability: Anthropological approaches and prospects. *Annual Review of Anthropology*, *39*, 401–415. https://doi.org/10.1146/annurev.anthro.012809.105045
[30] Otto, S., & Kaiser, F. G. (2014). Ecological behavior across the lifespan: Why environmentalism increases as people grow older. *Journal of Environmental Psychology*, *40*, 331–338. https://doi.org/10.1016/j.jenvp.2014.08.004
[31] Preacher, K. J., & Hayes, A. F. (2008). Asymptotic and resampling strategies for assessing and comparing indirect effects in multiple mediator models. *Behavior Research Methods*, *40*(3), 879–891. https://doi.org/10.3758/BRM.40.3.879
[32] Pretty, J., Adams, B., Berkes, F., De Athayde, S. F., Dudley, N., Hunn, E., … Pilgrim, S. (2009). The intersections of biological diversity and cultural diversity: Towards integration. *Conservation and Society*, *7*(2), 100–112. https://doi.org/10.4103/0972-4923.58642
[33] Priadi, A., Fatria, E., Sarkawi, D., & Oktaviani, A. (2018). *Environmental citizenship behavior ( the effect of environmental sensitivity , knowledge of ecology , personal investment in environmental issue , locus of control towards s tudents ' environmental citizenship behavior )*. 08002, 1–6.
[34] Recha, C. (2019). Regional Variations and Conditions for Agriculture in Kenya. *Current Politics and Economics of Africa*, *12*(1), 87.
[35] Saitabau, H., & Nairobi-kenya, P. O. B. (2014). *IMPACTS OF CLIMATE CHANGE ON THE LIVELIHOODS OF LOITA MAASAI PASTORAL COMMUNITY AND RELATED INDIGENOUS KNOWLEDGE ON ADAPTATION AND MITIGATION By Table of Contents*. 1–35.
[36] Saleem, M. A., Eagle, L., & Low, D. (2018). Market

segmentation based on eco-socially conscious consumers' behavioral intentions: Evidence from an emerging economy. In *Journal of Cleaner Production* (Vol. 193). https://doi.org/10.1016/j.jclepro.2018.05.067

[37] Services, M., Health, P., & Macro, I. C. F. (2010). *Kenya Service Provision Assessment Survey 2010 [SPA17]*.

[38] Snelgar, R. S. (2006). Egoistic, altruistic, and biospheric environmental concerns: Measurement and structure. *Journal of Environmental Psychology*, *26*(2), 87–99. https://doi.org/10.1016/j.jenvp.2006.06.003

[39] Sponarski, C. C., Vaske, J. J., & Bath, A. J. (2015). Attitudinal differences among residents, park staff, and visitors toward coyotes in cape breton highlands national park of Canada. *Society and Natural Resources*, *28*(7), 720–732. https://doi.org/10.1080/08941920.2015.1014595

[40] Steg, L. (2007). *VALUE ORIENTATIONS AND ENVIRONMENTAL BELIEFS IN FIVE COUNTRIES Validity of an Instrument to Measure Egoistic , Altruistic and Biospheric Value Orientations*. *38*(3), 318–332. https://doi.org/10.1177/0022022107300278

[41] Stern, P. C., Dietz, T., Abel, T., Guagnano, G. A., & Kalof, L. (1999). A value-belief-norm theory of support for social movements: The case of environmentalism. *Human Ecology Review*, *6*(2), 81–97.

[42] Stern, Paul C. (1995). *Values , Beliefs , and Proenvironmental Action : Attitude Formation Toward Emergent Attitude Objects1*.

[43] Stern, Paul C. (2000). *Toward a Coherent Theory of Environmentally Significant Behavior*. *56*(3), 407–424.

[44] Stern, Paul C, & Dietz, T. (1994). *The Value Basis of Environmental Concern*. *50*, 65–84.

[45] Taylor, B. M., & Van Grieken, M. (2015). Local institutions and farmer participation in agri-environmental schemes. *Journal of Rural Studies*, *37*, 10–19. https://doi.org/10.1016/j.jrurstud.2014.11.011

[46] Tidwell, T. (2010). A new environment for land and resource management: rising to the challenge. Andrus Conference, Boise, Idaho. *The Open Ecology Journal*, (7), 9–31. https://doi.org/10.2174/1874213001407010009

[47] Turaga, R. M. R., Howarth, R. B., & Borsuk, M. E. (2010). Pro-environmental behavior: Rational choice meets moral motivation. *Annals of the New York Academy of Sciences*, *1185*, 211–224. https://doi.org/10.1111/j.1749-6632.2009.05163.x

[48] Van Herzele, A., Gobin, A., Van Gossum, P., Acosta, L., Waas, T., Dendoncker, N., & Henry de Frahan, B. (2013). Effort for money? Farmers' rationale for participation in agri-environment measures with different implementation complexity. *Journal of Environmental Management*, *131*, 110–120. https://doi.org/10.1016/j.jenvman.2013.09.030

[49] Wolf, J. (1958). Über das Vorkommen von Chlorogensäure und inren Isomeren in Blättern von Steinobstbäumen. *Die Naturwissenschaften*, *45*(6), 130–131. https://doi.org/10.1007/BF00640994

[50] Zhao, H. H., Gao, Q., Wu, Y. P., Wang, Y., & Zhu, X. D. (2014). What affects green consumer behavior in China? A case study from Qingdao. *Journal of Cleaner Production*, *63*, 143–151. https://doi.org/10.1016/j.jclepro.2013.05.021